\documentclass[longauth]{aa}
\usepackage{graphicx}
\usepackage{txfonts}
\usepackage{natbib}
\usepackage{float}
\usepackage[colorlinks=true,citecolor=blue]{hyperref}
\usepackage{orcidlink}
\usepackage{xspace}
\usepackage{subfig}
\usepackage{lscape} 
\usepackage{ragged2e}
\usepackage[symbol]{footmisc}
\usepackage{caption}        
\usepackage{float}          
\usepackage[export]{adjustbox} 

\usepackage[export]{adjustbox}
\usepackage{sidecap}
\usepackage{mcaption}
\usepackage{placeins}
\sidecaptionvpos{figure}{c}

\bibpunct{(}{)}{;}{a}{}{,}

\newcommand{\MJup}{\ensuremath{M_{\mathrm{Jup}}}\xspace}

\newcommand{\MSun}{\ensuremath{M_{\odot}}\xspace}

\def\hr8799{{HR\,8799}}

\newcommand{\as}{\hbox{$^{\prime\prime}$}\xspace}

\begin{document}

   \title{Astrometric and photometric characterization of $\eta$ Tel B combining two decades of observations\thanks{This publication made use of data from the ESO programs 095.C-0298(A), 097.C-0394(A), 198.C-0209(H), 1100.C-0481(G)}}

\author{P. H. Nogueira\inst{1,2}\orcidlink{0000-0001-8450-3606}, C. Lazzoni\inst{3,2}\orcidlink{0000-0001-7819-9003}, A. Zurlo\inst{1,2}\orcidlink{0000-0002-5903-8316}, T. Bhowmik\inst{1,2}\orcidlink{0000-0002-4314-9070}, C. Donoso-Oliva\inst{4,2}\orcidlink{0009-0006-3064-8002}, S. Desidera\inst{5}\orcidlink{0000-0001-8613-2589}, J. Milli\inst{6}\orcidlink{0000-0001-9325-2511}, S. P\'erez\inst{7,8,2}\orcidlink{0000-0003-2953-755X
}, P. Delorme\inst{6}\orcidlink{0000-0002-2279-410X}, A. Fernadez\inst{4,2}\orcidlink{0009-0003-1316-6771}, M. Langlois\inst{5}\orcidlink{0000-0003-3574-9903}, S. Petrus\inst{1,2,7,8}\orcidlink{0000-0003-0331-3654}, G. Cabrera-Vives\inst{4,2}\orcidlink{0000-0002-2720-7218}, G. Chauvin\inst{9}\orcidlink{0000-0003-4022-8598}}
         
   \institute{\inst{1} Instituto de Estudios Astrof\'isicos, Facultad de Ingenier\'ia y Ciencias, Universidad Diego Portales, Av. Ej\'ercito 441, Santiago, Chile \\
\inst{2} Millennium Nucleus on Young Exoplanets and their Moons (YEMS), Chile \\
             \email{nogueirapedro404@gmail.com}\\
\inst{3} Department of Physics and Astronomy, University of Exeter, Stocker Road, Exeter, EX4 4QL, UK\\
\inst{4} Department of Computer Science, Universidad de Concepción, Concepción 4070386, Chile \\
\inst{5} INAF-Osservatorio Astronomico di Padova, Vicolo dell'Osservatorio 5, Padova, Italy, 35122-I \\
           \inst{6} Univ. Grenoble Alpes, CNRS, IPAG, 38000, Grenoble, France \\
              \inst{7} Departamento de Física, Universidad de Santiago de Chile, Av. Victor Jara 3659, Santiago, Chile \\
              \inst{8} Center for Interdisciplinary Research in Astrophysics and Space Science (CIRAS), Universidad de Santiago de Chile\\
 \inst{9} Laboratoire Lagrange, UMR7293, Université Côte d'Azur, CNRS, Observatoire de la Côte d'Azur, Boulevard de l'Observatoire, 06304, Nice, France  \\
}


 
\abstract{$\eta$ Tel is an 18 Myr system composed of a 2.09 M$_{\odot}$ A-type star with an M7-M8 brown dwarf companion, $\eta$ Tel B. The two objects have a projected separation of 4$\farcs$2 ($\sim$208 au). This system has been targeted by high-contrast imaging campaigns for over 20 years, facilitating its orbital and photometric characterization. The companion, $\eta$ Tel B, both bright and on a wide orbit, is an ideal candidate for a detailed examination of its position and the characterization of its atmosphere.} 
{To explore the orbital parameters of $\eta$ Tel B, measure its contrast, and investigate its close surroundings, we analyzed three new SPHERE/IRDIS coronagraphic observations. Our objectives are to investigate the possibility of a circumplanetary disk or a close companion around $\eta$ Tel B, and characterize its orbit by combining this new data set with archival data acquired in the past two decades.}
{The IRDIS data are reduced with state-of-the-art algorithms to achieve a contrast with respect to the star of 1.0$\times 10^{-5}$ at the location of the companion. Using the NEGative Fake Companion technique (NEGFC), we measure the astrometric positions and flux of $\eta$ Tel B for the three IRDIS epochs. Together with the measurements presented in the literature, the baseline of the astrometric follow-up is 19 years.}
  {We calculate a contrast for the companion of 6.8 magnitudes in the H band. The separation and position angle measured are 4$\farcs$218 and 167.3 degrees, respectively. The astrometric positions of the companions are calculated with an uncertainty of 4 milliarcseconds (mas) in separation and 0.2 degrees in position angle. These are the smallest astrometrical uncertainties of $\eta$ Tel B obtained so far. The orbital parameters are estimated using the Orvara code, including all available epochs. The orbital analysis is performed taking into account the Gaia-Hipparcos acceleration of the system. Suppressing its point spread function (PSF), we have produced contrast curves centered on the brown dwarf in order to constrain our detection capabilities for a disk or companions around it.}
{After considering only orbits that could not disrupt the outer debris disk around $\eta$ Tel A, our orbital analysis reveals a low eccentric orbit (e $\sim$ 0.34) with an inclination of 81.9 degrees (nearly edge-on) and a semi-major axis of 218 au. Furthermore, we determine the mass of $\eta$ Tel B to be 48 \MJup, consistent with previous calculations from the literature based on evolutionary models. Finally, we do not detect any significant residual pointing to the presence of a satellite or a disk around the brown dwarf. The retrieved detection limits allow us to discard massive objects around $\eta$ Tel B with masses down to 1.6 \MJup at a separation of 33 au.}

\titlerunning{ Characterization of the $\eta$\,Tel system }
\authorrunning{P. H. Nogueira et al.}
   \maketitle 
   
%

\section{Introduction}

Since the first discovery of a planet around a main-sequence star \citep{mayorequeloz1995}, the techniques and instruments for discovering and characterizing substellar objects have advanced exponentially. Today, more than 5500 exoplanets have been confirmed\footnote{Information extracted from the NASA Exoplanet archive database \url{https://exoplanetarchive.ipac.caltech.edu/} \citep{akeson2013}.}, and more than 19,000 candidates of ultracool dwarfs (spectral type later than M7) are known \citep{dalPonte2023}. However, little is known about where, when, and how substellar objects form. Thus, investigating newborn substellar objects in a stellar system is essential to address this knowledge gap, since they are expected to retain signatures of their formation pathway through the system architecture (e.g., \citealp{2004ApJ...609.1045M,2014MNRAS.443.2541P,2014MNRAS.438L..31G,2019MNRAS.484.1926D,2020AJ....159...63B,2023AJ....166...48D}) and from their atmosphere (e.g., \citealp{2007prpl.conf..733M,2011ApJ...743L..16O,2014ApJ...790..133D,2015A&A...577A..42B,2018ApJ...853..192C,2023AJ....166..198Z,2024ApJ...966L..11P}).

Particularly, the high-contrast imaging technique (hereafter called HCI) is a capable tool to discover and characterize young giant planets and brown dwarfs. HCI is sensitive to recently formed objects that conserve some formation heat. Indeed, substellar objects have been discovered using HCI techniques in the past decade (e.g. 2MASSWJ 1207334-393254: \citealp{chauvin2004}; $\beta$ Pic b: \citealp{lagrange2010}; HD 95086 b: \citealp{rameau2013discovery}; PDS 70 b: \citealp{keppler2018}; YSES 2b: \citealp{bohn2021}). More recently, with the release of the Gaia catalogs and an update on astrometric precision, HCI has also been used to follow up on stars that showed anomalous accelerations (difference between their long-term HIPPARCOS-Gaia and short-term Gaia proper motion vectors), pointing towards the presence of companions in the system. One remarkable example is the discovery of AF Lep b through proper motion anomalies and HCI \citep{mesa,2023A&A...672A..94D,2023ApJ...950L..19F}.

Near-infrared substellar companions observed by High-Contrast Imaging (HCI) may offer valuable multi-wavelength and multi-analysis data for the detection of circumplanetary disks (CPDs) or satellites in their surroundings. For instance, \citet{perez2019} analyzed ALMA band 6 observations featuring directly imaged companions, establishing upper limits on CPD detectabilities. Similarly, using SPHERE near-infrared (NIR) images, \citet{lazzoni2020} sets upper limits for satellite detections around substellar companions. Notably, \citet{lazzoni2020} also identifies a potential satellite candidate around DH Tau B, although confirmation is pending. Subsequently, \citet{lazzoni2022} provides a more robust analysis regarding the detectability of satellites through all standard exoplanet discovery techniques. Additionally, \citet{ruffio2023} proposes directing efforts towards detecting potential candidates through radial velocity monitoring, a method that can be implemented by monitoring the companions with high-resolution spectroscopy. Another technique to look for satellites is spectroastrometry, which consists of the fine measurement of any deviation of the position of the center of light \citep{2015ApJ...812....5A}. If an integral field spectrograph targets an unresolved planet-satellite system, it is expected that the center of the light shifts position depending on the wavelength. The movement of the centroid of the PSF would reveal if a satellite is present.

One system that comes to attention based on its age, being targetable from HCI and satellite analysis, and its astrometric follow-up is $\eta$ Telescopii (hereafter called $\eta$ Tel). $\eta$ Tel is part of the $\beta$ Pic moving group, with an estimated age of 18 Myr \citet{2020A&A...642A.179M} and a distance of 49.5 pc \citep{gaiadr3}. It is composed of an A0V, 2.09 M$_{\odot}$ primary star ($\eta$ Tel A; \citealp{houk1975, desidera2021,desidera2021catalogue}) and an M7-8, brown dwarf companion ($\eta$ Tel B) at $\sim$4$\farcs$2 separation and position angle of $\sim$169$^{\circ}$ (\citealp{lowrance2000, guenther2001}, and references therein). $\eta$ Tel A presents a likely warm debris belt at 4 au (unresolved, only inferred from the SED) and an edge-on cold debris belt at 24 au discovered via infrared excess \citep{backman1993,mannings1998}. The outer debris disk was later resolved with T-ReCS (Thermal-Region Camera Spectrograph) on Gemini South \citep{2009A&A...493..299S}. It also shows a radiatively driven debris disk wind, consistent with C/O solar ratio \citep{youngblood2021}. 
$\eta$ Tel B is a bright substellar companion with a contrast of 6.7 magnitudes in VLT/NACO H band \citep{neuhauser2011}, or 11.85 in apparent magnitude. Its astrometrical points and orbital constraints were first compiled and analyzed in \cite{neuhauser2011}, which used 11 years of imaging data (1998-2009). One more NACO observation was presented in \cite{rameau2013}, which broadened the time baseline to 2011.

In this paper, we present a characterization of the companion, including astrometrical and photometrical follow-up and orbital constraint analysis. We present three new epochs from the Spectro-Polarimetric High-contrast Exoplanet REsearch (SPHERE) instrument \citep{beuzit2019}. With the addition of the new SPHERE data and a baseline of observations from 1998 to 2017, we present the most recent and complete orbital characterization of the system. Additionally, we performed an analysis of the surroundings of the companion $\eta$ Tel B to constrain the possible presence of features such as satellites or circumplanetary disks. 

The manuscript is structured in the following order. The observations and data reduction are described in Section \ref{sec:obs}. The photometric and astrometric measurements of the system of the new SPHERE observations are reported in Section \ref{sec:astrometry}. The orbital fitting analysis, taking into account the new data and the literature can be found in Section \ref{sec:orbfit}. The description of the study of the close vicinity of the substellar companion and the contrast curves around $\eta$ Tel B are presented in Section \ref{sec:sathunt}. Final remarks and conclusions are presented in Section \ref{sec:sum}. 

\section{Observations and data reduction} 
\label{sec:obs}
\subsection{Observations}
We present new SPHERE/IRDIS coronagraphic data of the system around $\eta$ Tel A. SPHERE is a VLT (Very Large Telescope) planet-finder instrument located at Paranal, Chile - UT3. It is an instrument dedicated to performing HCI. SPHERE is composed of four main scientific parts: the Common Path and Infrastructure (CPI), which includes an extreme adaptive optics system (SAXO, \citealp{fusco2006,petit2014}) and coronagraph systems; the Infrared Dual-band Imager and Spectrograph (IRDIS, \citealp{dohlen2008}) with a pixel scale of 12.25 mas and field of view (FOV) of 11$\times$12.5 arcsec; the integral field spectrograph (IFS, \citealp{claudi2008}) with FOV of 1\farcs73$\times$1\farcs73; and the Zurich Imaging Polarimeter (ZIMPOL, \citealp{schmid2018}), the visible light imager and polarimeter of SPHERE. Both IRDIS and IFS belong to the NIR branch and can be used concomitantly if required during observations. Given the separation of $\eta$ Tel B and its brightness in the near-infrared, IRDIS observations are the most suitable for the purpose of our paper. 

The observations were performed with IRDIS on three different nights. Since the stellar companion is outside the field of view (FoV) of the integral field spectrograph (IFS) instrument, we do not report its data in this manuscript. The first observation (program 095.C-0298(A); PI Beuzit) was taken on 2015-05-05, generally under average conditions, with variable seeing ranging between 1 and 2 arcsecs (this epoch had photometric sky transparency). The second observation (program 097.C-0394(A); PI Milli) was taken under good conditions on 2016-06-15, and the seeing was mostly stable during the period of observation, varying between 1-1.4 arcsecs (this epoch had the sky transparency declared as thin). This program was taken as part of the SPHERE High-Angular Resolution Debris Disks Survey (SHARDDS; \citealp{dahlqvist2022}), designed to image circumstellar disks around bright stars, closer than 100 pc. The third night (program 198.C-0209(H); PI Beuzit) was on 2017-06-15, with a slightly higher value of the seeing and poorer conditions (``thin'' sky transparency). An apodized Lyot coronagraph (N\_ALC\_YJH\_S; inner working angle $\sim$0$\farcs$15) was used on all three nights. A dual-band filter H2H3 was set ($\lambda$= 1.593 and 1.667 $\mu$m for H2 and H3, respectively; $\Delta$$\lambda$= $\sim$0.053 $\mu$m for both filters) on the first and third nights, while on the second night, a broadband filter H ($\lambda$= 1.625 $\mu$m; $\Delta$$\lambda$= 0.29 $\mu$m) was used instead. A fourth IRDIS sequence was taken (2018-05-08; ID: 1100.C-0481(G); PI: Beuzit), but it was discarded from our analysis due to bad observing conditions. The observations and their specifications are summarized in Table~\ref{table_obs}.

\begin{table*}[!htb]
\centering
\caption{List of the SPHERE/IRDIS $\eta$ Telescopii observations used in this work.}
\begin{tabular}{ccccccccc}
\hline
\hline
Date (UT)  & ESO ID - PI            & IRDIS filter & DIT $\times$ NDIT & $\Delta$PA & Seeing ["]& Avg. coherence time [s]\\
\hline
2015-05-05 & 095.C-0298(A)- Beuzit & DB\_H2H3     & 32 $\times$ 8               & 46.83  & Mostly 0.9-1.5$^a$  &   0.0011    \\
2016-06-15 & 097.C-0394(A)- Milli  & BB\_H        & 8 $\times$ 8                & 18.17  &  1-1.4 &   0.0024       \\
2017-06-15 & 198.C-0209(H)- Beuzit & DB\_H2H3     & 32 $\times$ 12              & 2.04   & 1.5-2  &   0.0023    \\
2018-05-08$^b$ & 1100.C-0481(G)- Beuzit &  --    & -- & --     & --  &   --   \\
\hline
\end{tabular}
\\
\Centering
\raggedright {\footnotesize $^a$A subtle increase in seeing above 1.5 was registered between 08h40m-9h00m UTC at that night.}
 \\
\raggedright {\footnotesize $^b$Discarded from the analysis for bad observing conditions.  }
\label{table_obs}
\end{table*}
\subsection{Data reduction}
The reduction of the three epochs was performed by the High Contrast Data Center pipeline\footnote{Formerly known as SPHERE Data Center or SPHERE DC.} (hereafter called as HC DC; \citealp{delorme2017}), which utilizes the Data Reduction and Handling software \citep[v0.15.0;][]{pavlov2008} and routines presented in \cite{2018A&A...615A..92G}. The process includes standard pre-reduction steps such as background subtraction, flat-fielding, and bad pixel correction. The frames are recentered using the SPHERE waffle pattern, followed by corrections for the anamorphism of the instrument and astrometric calibration (pixel scale and True north correction), as described in \citet{maire2016}. The final products comprise a master cube that contains all frames, the position angle (PA) values of each frame, and an off-axis PSF reference cube. The off-axis PSF is an unsaturated image of the central star taken before and after the coronagraphic sequence for flux calibration.  

\begin{figure}[!h]
\begin{center}
    \includegraphics[width=0.48\textwidth]{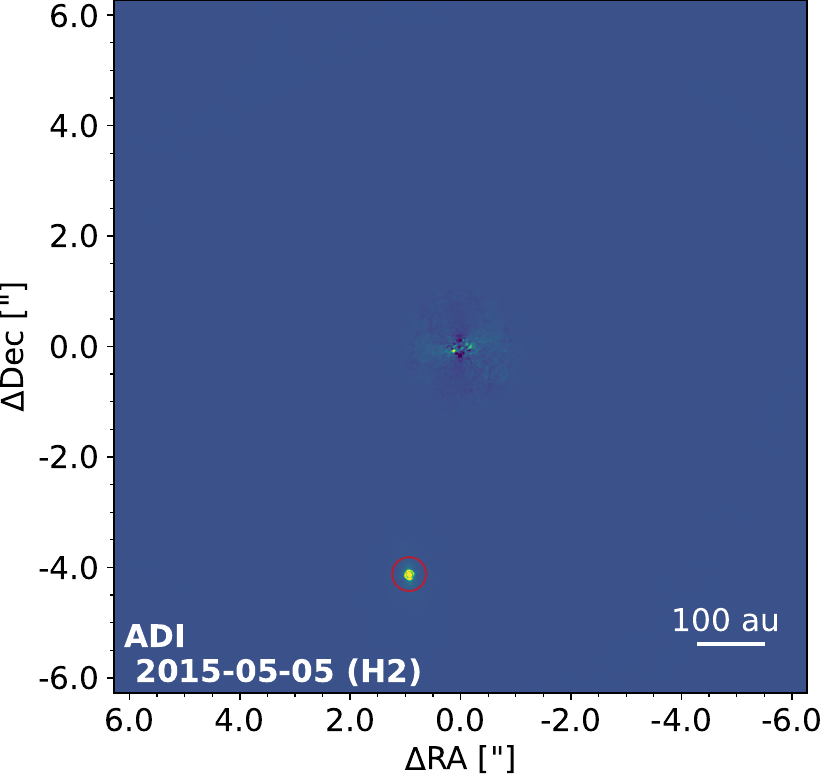}
    \caption{ADI-processed from IRDIS coronagraphic sequence (epoch: 2015-05-05; filter: H2) for the $\eta$ Tel system. The red circle marks the position of $\eta$ Tel B. The central star is masked under the coronagraph at the center of the image. North is up, East is left.}
    \label{adis}
\end{center}
\end{figure}

We made use of the VIP code (Vortex Image Processing, \citealp{vip2017}; v.1.5.1) to reject the bad frames in each master cube, considering only frames with a Spearman correlation above 0.85 compared to the first frame taken under good conditions. After this step, we created post-processing images, employing the Angular Differential Imaging \citep[ADI;][]{marois2006} and principal component analysis (PCA; \citealp{soummer2012, amaraequanz2012}) with varying numbers of principal components.
Since the PCA introduces deep over-subtraction due to the FoV rotation even for a few components, we favored the classical ADI images for our analysis. The results of this post-processing technique are presented in Fig. \ref{adis}.

\section{Methodology and results}

\subsection{Astrometry and photometry}
\label{sec:astrometry}

\subsubsection{The NEGFC technique}
\label{sec:astrometry_negfc}

To precisely determine the position and flux of $\eta$ Tel B, we employed the NEGative Fake Companion Technique (NEGFC), as described in studies such as \citet{lagrange2010} and \citet{zurlo2014}. This technique involves modeling the target source by introducing a negative model of the instrumental PSF into the pre-processed data. The procedure aims to minimize residuals in the final image, adjusting the flux and position of the model to align with the source's properties (astrometry and photometry). We implemented the NEGFC using the VIP package.

For this purpose, we adapted the {\tt single\_framebyframe} routine proposed by \citet{lazzoni2020}. This routine provides estimates of separation, position angle, and photometry for the companion in each frame from the coronagraphic sequence. $\eta$ Tel B's brightness (signal-to-noise ratio $\geq$100) and relative distance from the speckle-dominated region are sufficient for this approach to be applicable.

As a model PSF, we utilized the off-axis image of the central star captured before and after the coronagraphic sequence. For each set of coordinates, a negative flux was introduced, and the set of positions and fluxes yielding the lowest residual (standard deviation) in each frame was selected. Consequently, for each night and filter, the routine's results were represented by the median values of the parameters measured across all frames.

\subsubsection{Measurement of the photometry}
\label{sec:phot}
\begin{figure}[!htb]
\begin{center}
    \includegraphics[width=0.5\textwidth]{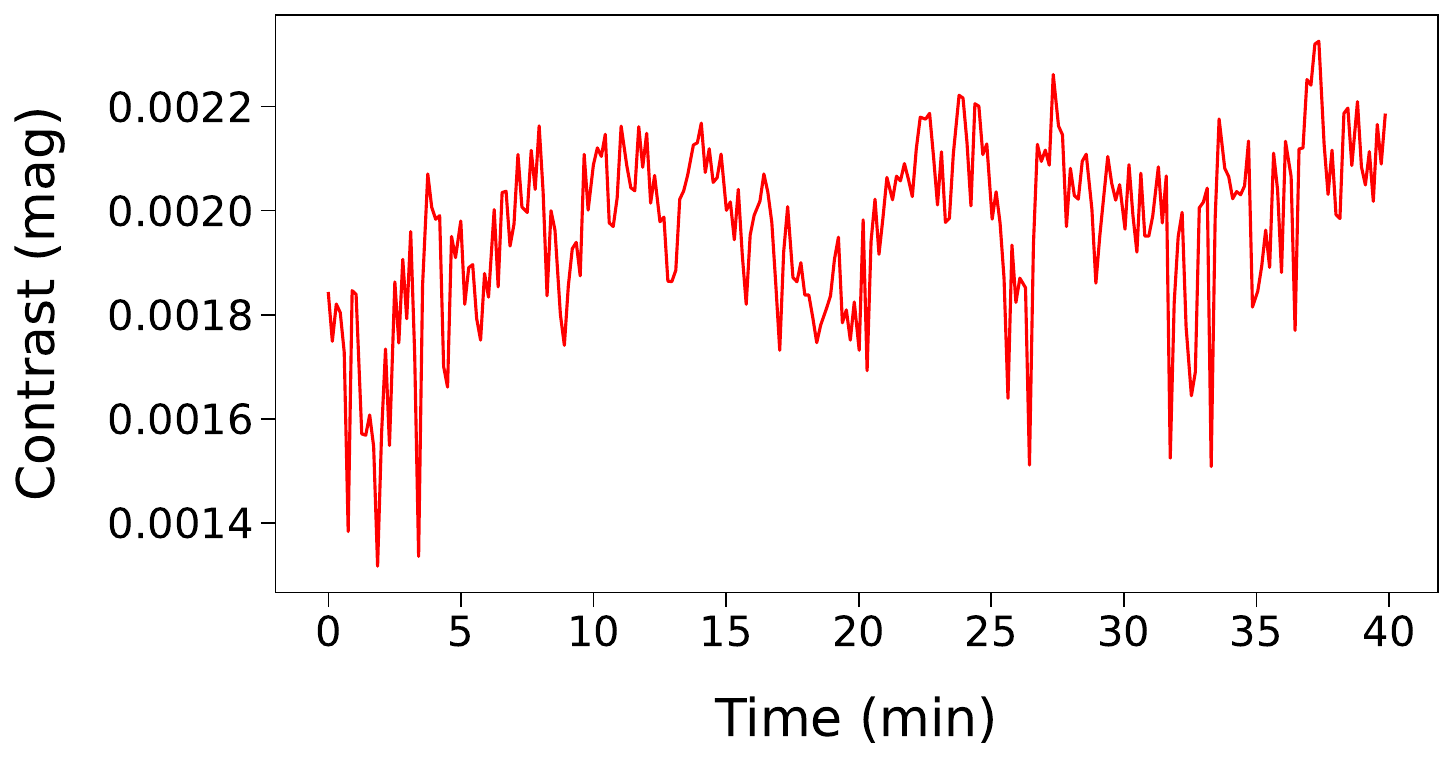}
    \caption{Contrast flux variations for the companion with respect to the star (filter: BB\_H).}
    \label{flux_variations}
\end{center}
\end{figure}

The contrast of the brown dwarf was calculated as the median of the NEGFC technique values on each frame of the coronographic sequence. Flux contrast uncertainties were derived from the standard deviation of the fluxes measured by the NEGFC technique. Fig. \ref{flux_variations} sets flux measurements per frame using BB\_H filter, showing how the values can vary. The contrast in flux and magnitude with respect to the central star for each filter is shown in Table \ref{table_phot}. The results are consistent with the measurements presented in the literature for the companion.

\begin{table}[!h]
\centering
\caption{IRDIS flux contrast in flux and magnitude with respect to the central star for $\eta$ Tel B.}
\begin{tabular}{cccc}
\hline
\hline
Epoch & Filter        & Contrast [e-3]                                & $\Delta$ magnitude           
\\
\hline

2015-05-05 & H2 & 1.52 $\pm$ 0.60   & 7.05 $\pm$ 0.45   \\
2015-05-05 & H3 & 1.92 $\pm$ 0.12   & 6.79 $\pm$ 0.07   \\

2016-06-15 & BB\_H & 2.00 $\pm$ 0.17 & 6.75 $\pm$ 0.09 \\

2017-06-15 & H2    & 1.53 $\pm$ 0.78 & 7.04 $\pm$ 0.61  \\
2017-06-15 & H3    & 1.92 $\pm$ 0.84 & 6.79 $\pm$ 0.51

\end{tabular}
\label{table_phot}
\end{table}

\begin{figure}[!h]
\begin{center}
    \includegraphics[width=0.47\textwidth]{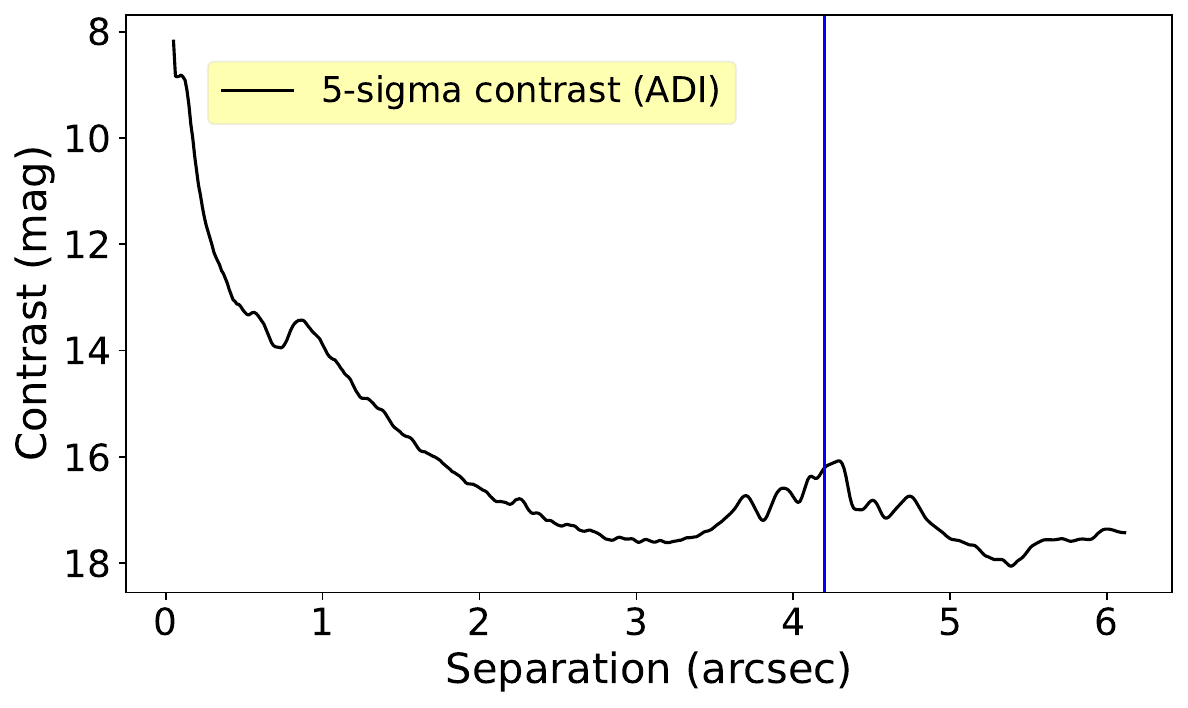}
    \caption{Contrast curve with respect to $\eta$ Tel A, using the datacube corresponding to the 2016-06-15 (BB\_H) observation. The vertical blue line corresponds to the position of the companion.}
    \label{contrastcurve_aroundstar}
\end{center}
\end{figure}

After using the NEGFC technique, the signal of the companion was removed in each frame of the datacubes, creating therefore empty datacubes. Following, a 5-sigma contrast curve with respect to and around $\eta$ Tel A was produced using the ADI processed data. The achieved contrast for epoch 2016-06-15 is shown in Fig. \ref{contrastcurve_aroundstar}.

\subsection{Measurement of the astrometric positions}
\label{astro_determination}

To robustly determine the separation and position angle at each epoch, we applied multiple methods while concurrently estimating the disparities between these algorithms. Specifically, we employed: 

\begin{itemize}
    \item the NEGFC method detailed in Section \ref{sec:astrometry_negfc};
    \item a 2D-Gaussian fitting encompassing the companion. We optimized Gaussian statistics using two stochastic algorithms: Adam \citep{kingma2014adam} and the Levenberg–Marquardt algorithm \citep{more2006levenberg};
    \item the peak intensity pixel location within a FWHM of the companion.
\end{itemize} 


All the algorithms were applied to each frame of the coronagraphic (frame-by-frame) sequence as well as to their median-collapsed reduction. When using the frame-by-frame method, it is important to note that a sequence of positions is obtained and it is necessary to reduce them by using the median. Finally, the astrometric points for each night and filter were determined by calculating the median among the outputs of the algorithms. Similarly, the uncertainties of the astrometrical fitting were calculated as the standard deviation across all results obtained from the various methods. We justify applying different methods as we observed that the final results of each method could differ by a maximum of 10 mas in separation and approximately 0.15 degrees in position angle. Therefore, we opted for implementing a median instead of an average to filter out values that may otherwise skew the results.

In addition to the uncertainties on the astrometric fitting of the companion, we have to consider other factors in the error budget, as previously stated by \citet{wertz2017}:
\begin{itemize}
    \item Instrumental calibration, where the most relevant errors come from the orientation of the True North, pupil offset, plate-scale, and anamorphism;
    \item Determination of the position of the central star behind the coronagraph;
    \item Systematic error due to residual speckles;
    \item Statistical error due to planet position determination.
\end{itemize}

Therefore, considering R as the final expression for radial separation in arcseconds and $\Theta$ as the final expression for position angle in degrees, we applied the following approximated equations:

\begin{equation}\label{eq:R}
    R = PS ( R_{*} \pm R_{spec} \pm r R_{AF} \pm r)\\ \text{and}
\end{equation}

\begin{equation}\label{eq:theta}
 \Theta =  \Theta_{*} \pm \Theta_{spec} \pm \Theta_{AF} \theta \pm \Theta_{PO} \pm \Theta_{TN} \pm \theta \\ \text{, }
\end{equation}

\noindent  
 where r is the radial distance and $\theta$ the position angle, $R_{*}$ and $\Theta_{*}$ the radial and azimuthal values related to stellar centering; $R_{spec}$ and $\Theta_{spec}$ the radial and azimuthal values related to speckle noise;  $R_{AF}$ and $\Theta_{AF}$ values related to the anamorphic factor expressed in percentage; TN related to the true north, PS to plate scale (\as/pixel) and PO to the pupil offset. All distances are measured in pixels and angles in degrees.

Consequently, we can use the Equations \ref{eq:R} and \ref{eq:theta} to propagate the errors:

\begin{equation}
    \sigma _{R}^{2}= PS^{2}[ \sigma _{R_{spec}}^{2} + \sigma _{R_{*}}^{2}  + r^{2}\sigma _{R_{AF}}^{2} + ( R_{AF} + 1 )^{2} \sigma _{r}^{2}] + \frac{R^{2}}{PS^{2}}\sigma_{PS}^{2} 
\end{equation}

\noindent and,

\begin{equation}
    \sigma _{\Theta}^{2}=\sigma _{\Theta_{spec}}^{2} + \sigma _{\Theta_{*}}^{2} +  \theta^{2} \sigma _{\Theta_{AF}}^{2} + \sigma _{PO}^{2} + \sigma _{TN}^{2} + (\Theta_{AF} + 1)^ {2}\sigma _{\theta}^{2} 
    \\ \text{. }
\end{equation}

The instrumental calibration uncertainties are determined through astrometric calibrations outlined in \citet{maire2016}, \citet{2021JATIS...7c5004M}, and the last version of the SPHERE manual, 18th release. Before July 2016, an issue with the synchronization between SPHERE and VLT internal clocks led to abnormal fluctuations in True North measurements. Consequently, for the initial two $\eta$ Tel observations, the True North uncertainties were extracted from Table 3 of \citet{maire2016}. For the third observation, we adopted a fixed value of 0.04$^{\circ}$, representing the stabilized uncertainty following calibrations and corrections.

The plate (pixel) scale uncertainties, were extracted from close-in-time coronagraphic observations, obtained with the SPHERE-SHINE GTO data, using the globular stellar cluster
47 Tuc as field reference (Table 7 of \citealt{2021JATIS...7c5004M}). 
The pupil offset uncertainty, derived from commissioning and guaranteed time observations, is 0.11 degrees. Distortion is predominantly influenced by a 0.60\% $\pm$ 0.02\% anamorphism between the horizontal and vertical axes of the detector. Since each frame has undergone correction by the HC DC, rescaling each image by 1.006 along the axis, the uncertainty of 0.02\% (for both $R_{AF}$ and $\Theta_{AF}$) was incorporated into the error budget analysis.

The radial stellar centering uncertainty per dithering is 1.2 mas, derived from observations of bright stars during commissioning runs \citep{zurlo2014, zurlo2016}. This value was then adjusted by dividing it by the square root of the number of frames per observation. Subsequently, the latter was translated into an uncertainty on position angle by division by the separation $r$ of the companion. 

\begin{figure*}[!htb]
\begin{center}
\includegraphics[width=0.98\textwidth]{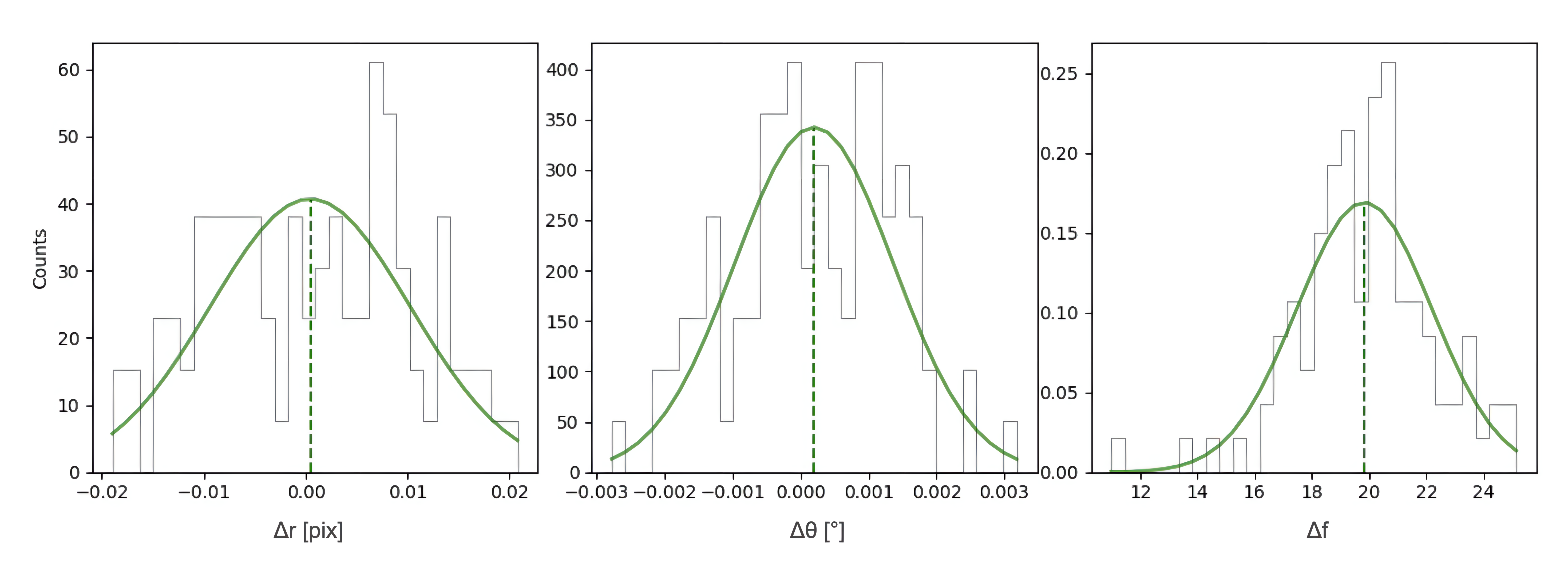}
    \caption{Speckle noise estimation for $\eta$ Tel B in the data set of 2015-05-05. The histograms illustrate the offsets between the true position and flux of a fake companion and its position and flux obtained from the NEGFC technique. The dashed lines correspond to the 1D Gaussian fit from which we determined the speckle noise.}
    \label{speckleuncert}
\end{center}
\end{figure*}

Uncertainties arising from speckles may persist even after ADI post-processing and have an impact on photometric and astrometric measurements \citep{guyon2012,wertz2017}. To address this, we employed the {\tt speckle\_noise\_uncertainty} function from VIP, injecting multiple simulated companions into companion-free cubes at the same radial distance and flux as the actual companion. The positions of these simulated companions were determined using Nelder-Mead optimization, and the values of separation and position angle were measured. By comparing the offsets between the input values and the estimations by the code, a distribution of parameters was generated. A Gaussian function was then fitted to the distribution, and the uncertainties in $R$ and $\Theta$ were estimated as the standard deviations of the fitting. In this instance, a total of 100 simulated companions, equally spaced azimuthally, were utilized. A similar methodology was employed by \citet{maire2015, wertz2017}. In Fig.~\ref{speckleuncert}, three histograms illustrate the distribution of separation, position angle, and flux of the companion observed on 2015-05-05. The separation and position angle values for the epoch are shown in Table~\ref{tab_framebyframe}, and the detailed uncertainties used to calculate the error budget are compiled in Table~\ref{tabuncert}. 

\begin{table}[!htb]
\centering
\caption{List of astrometric positions of $\eta$ Tel B obtained from the IRDIS observations. The error budget is listed in Table~\ref{tabuncert}.}
\begin{tabular}{cccccc}
\hline
Epoch & Filter$^a$ & Sep. (\as) & PA ($^{\circ}$) \\

\hline
\hline
2015-05-05  & H3 & 4.215$\pm$0.004                                                     &         167.326$\pm$0.197                                                               &                                                        \\
2016-06-15  &  BB\_H & 4.218$\pm$0.004                                                  &                167.260$\pm$0.130                                                        &                                                        \\
2017-06-15 &  H2 & 4.218$\pm$0.004                                                 &             167.346$\pm$0.142                                                           &                                                       
\end{tabular}
\label{tab_framebyframe}
\raggedright \footnotesize{  $^{a}$ The table only presents values related to the filters where the lowest uncertainties, retrieved from NEGFC, on each epoch, were achieved.}
\end{table}




\begin{table*}[!htb]
\caption{Uncertainties used to calculate the error budget for the astrometry of the companion. The values correspond to errors associated with individual frames.}
\centering
\begin{tabular}{c|ccccccccccc}
\hline
Epoch      & \begin{tabular}[c]{@{}c@{}}$\sigma_{r}$\\ (pixels)\end{tabular} & \begin{tabular}[c]{@{}c@{}}$\sigma_{R_{spec}}$\\ (pixels)\end{tabular} & \begin{tabular}[c]{@{}c@{}}$\sigma_{R_{*}}$\\ (pixels)\end{tabular} & \begin{tabular}[c]{@{}c@{}}$\sigma_{R_{AF}}$\\ (\%)\end{tabular} & \begin{tabular}[c]{@{}c@{}}$\sigma_{PS}$\\ (mas/pixel)\end{tabular} & \begin{tabular}[c]{@{}c@{}}$\sigma_{\theta}$\\ ($^{\circ}$)\end{tabular} & \begin{tabular}[c]{@{}c@{}}$\sigma_{\Theta_{spec}}$\\ ($^{\circ}$)\end{tabular} & \begin{tabular}[c]{@{}c@{}}$\sigma_{\Theta_{*}}$\\ ($^{\circ}$)\end{tabular} & \begin{tabular}[c]{@{}c@{}}$\sigma_{\Theta_{AF}}$\\ (\%)\end{tabular} & \begin{tabular}[c]{@{}c@{}}$\sigma_{PO}$\\ ($^{\circ}$)\end{tabular} & \begin{tabular}[c]{@{}c@{}}$\sigma_{TN}$\\ ($^{\circ}$)\end{tabular} \\
\hline
\hline
2015-05-05 & 0.297                                                          & 0.003                                                                      & 0.098                                                               & 0.02                                                             & 0.01                                                               & 0.068                                                                 & 0.0005                                                                            & 0.016                                                                         & 0.02                                                                  & 0.11                                                             & 0.145                                                             \\
2016-06-15 & 0.065                                                          & 0.001                                                                      & 0.098                                                               & 0.02                                                             & 0.01                                                               & 0.012                                                                 & 0.0001                                                                            & 0.016                                                                         & 0.02                                                                  & 0.11                                                             & 0.060                                                              \\
2017-06-15 & 0.245                                                          & 0.010                                                                      & 0.098                                                               & 0.02                                                             & 0.01                                                               & 0.072                                                                & 0.0015                                                                            & 0.016                                                                         & 0.02                                                                  & 0.11                                                             & 0.040                                                             
\end{tabular}
\\
\raggedright \footnotesize{The table only presents values related to the filters where the lowest uncertainties on each epoch were achieved: 2015-05-05 (H3); 2016-06-15 (BB\_H), and 2017-06-15 (H2), thus referring to the filters also listed on Table \ref{tab_framebyframe}.}
\label{tabuncert}
\end{table*}






\subsection{Orbital fitting analysis}
\label{sec:orbfit}

\begin{SCfigure*}
\includegraphics[width=0.75\textwidth]{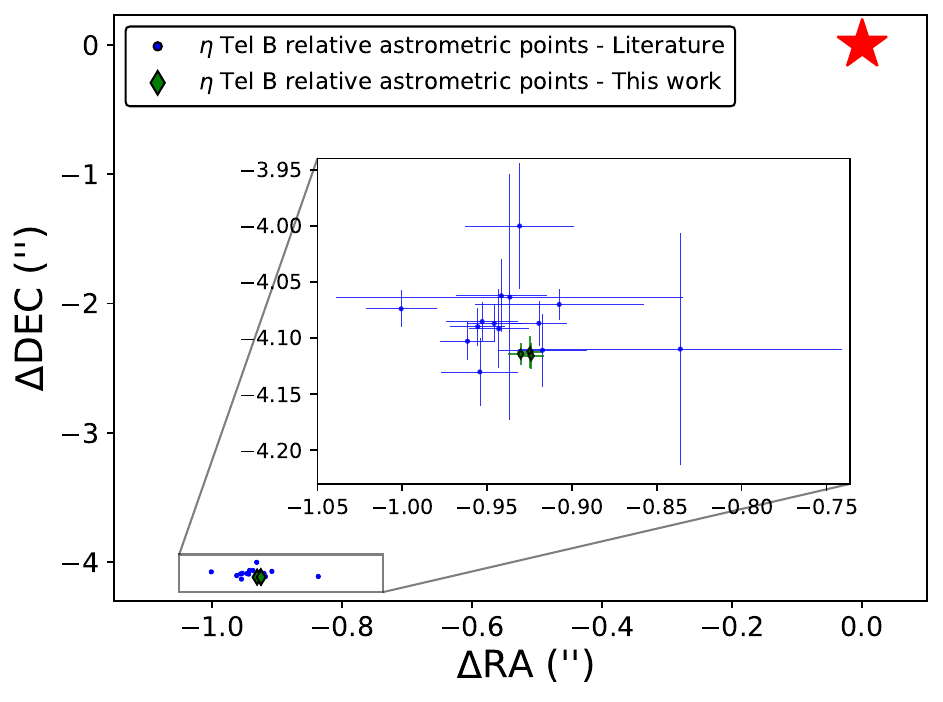}
    \caption{Relative astrometry for $\eta$ Tel B. The red star marks the position of $\eta$ Tel A. The blue-filled circles represent astrometric points extracted from the literature. The green-filled diamonds represent the astrometric points from this work. The astrometric point from epoch 2000.378 was not included due to its high uncertainties.}
    \label{relat_astrom}
\end{SCfigure*}

The system around $\eta$ Tel A was observed with the high-contrast imaging technique for the last two decades. The favorable contrast of the brown dwarf companion and the wide separation between the two objects make the system an ideal target for HCI. The astrometric follow-up of the brown dwarf companion counts 18 epochs of observation spanning almost 20 yr. For the orbit analysis, we included all the astrometrical points presented in the literature, as well as the ones obtained from the new analysis. The complete list is shown in Table~\ref{table_relast} and their positions with respect to the central star are represented in Figure~\ref{relat_astrom}.

\begin{table}[h!]
\centering
\caption{Astrometric positions of $\eta$ Tel B from the literature and our analysis included in the Orvara analysis.}
\begin{tabular}{llll}
Date (yr) & Separation (\as) & PA ($^{\circ}$) & Ref. \\
\hline
\hline
1998.492  & 4.170$\pm$0.033       & 166.95$\pm$0.36  &    N11\\
2000.307  & 4.107$\pm$0.057       & 166.90$\pm$0.42  &   N11\\
2000.378  & 4.310$\pm$0.270       & 165.80$\pm$6.70  &    G01 \\
2004.329  & 4.189$\pm$0.020       & 167.32$\pm$0.22  &   N11 \\
2004.329  & 4.200$\pm$0.017       & 166.85$\pm$0.22  &   N11\\
2004.329  & 4.199$\pm$0.036       & 167.02$\pm$0.22  &    N11\\
2004.329  & 4.195$\pm$0.017       & 166.97$\pm$0.22  &    N11\\
2006.431  & 4.170$\pm$0.110       & 167.02$\pm$1.40  &   G08 \\
2007.753        & 4.212$\pm$0.033       & 167.42$\pm$0.35  &  N11\\
2008.312   & 4.214$\pm$0.017       & 166.81$\pm$0.22  &    N11\\
2008.599   & 4.195$\pm$0.017       & 166.87$\pm$0.29  &    N11\\
2008.599   & 4.194$\pm$0.016       & 166.20$\pm$0.29  &   N11\\
2009.351        & 4.239$\pm$0.104       & 168.50$\pm$1.30  &   N11 \\
2009.496  & 4.199$\pm$0.031       & 166.99$\pm$0.30  &    N11 \\
2011.576  & 4.170$\pm$0.009       & 167.43$\pm$0.70  &    R13\\
2015.341  & 4.215$\pm$0.004       & 167.33$\pm$0.20  &    $^a$\\
2016.454  & 4.218$\pm$0.004      & 167.26$\pm$0.13  &    $^a$\\
2017.452   & 4.218$\pm$0.004       & 167.35$\pm$0.14   &  $^a$\\
\hline
\end{tabular}
\justifying
\\
 \footnotesize N11: \citet{neuhauser2011}; G01: \citet{guenther2001}; \\ G08: \citet{geissler2008}; R13: \citet{rameau2013}; \\ $^a$: This work. \hspace*{-\labelsep}
\label{table_relast}
\end{table}

We employed the Orvara code \citep[Orbits from Radial Velocity, Absolute, and/or Relative Astrometry;][]{2021AJ....162..186B} to perform the orbital fitting of the system. Orvara accepts input information on the acceleration of the central star from the Hipparcos vs Gaia early data-release 3 \citep[EDR3;][]{gaiadr3} catalogs.

The orbital elements and masses of both the host and the companion were computed by Orvara using a parallel tempered Markov Chain Monte Carlo (MCMC) ensemble sampler, {\tt ptemcee} \citep{vousden2016}, a variant of {\tt emcee} \citep{foremanmackey2013}. Parallel tempering enhances the exploration of complex parameter spaces by simultaneously running multiple chains at different temperatures, allowing for more efficient sampling and improved convergence compared to traditional MCMC methods. In a simulation, different temperatures refer to variations in the parameter that control the acceptance of proposed moves in the MCMC algorithm. Higher temperatures encourage more exploration by accepting moves that might increase energy or objective function values, allowing the algorithm to escape local optima and explore a broader solution space. On the other hand, lower temperatures favor exploitation, focusing on refining solutions and improving the chance of finding the global optimum. Adjusting temperatures during the simulation influences the balance between exploration and exploitation, shaping the algorithm's behavior throughout the MCMC process. The MCMC simulation utilized 10 temperatures, 500 walkers, and 10$^{6}$ steps for each chain. The simulation outputs a point every 1000 steps.

Priors for the MCMC include distributions of the masses of celestial objects, parallax, and proper motion of the system. Additionally, the initial orbital elements distribution for the companions (semi-major axis: \(a\); eccentricity: \(e\); argument of the pericenter: \(\omega\); inclination: \(i\); longitude of the ascending node: \(\Omega\); and longitude at reference epoch: \(l\)) can also be incorporated. We used as priors the common proper motion extracted from \citet{kervellacatalogue,kervella2022}, the parallax from Gaia DR3 \citep{gaiadr3}, the primary mass from \citet{desidera2021,desidera2021catalogue}, and the companion ($\eta$ Tel B) mass with loosened constraints on its uncertainties from \citet{lazzoni2020}. Given the low orbital coverage from relative astrometric points, we just set a mean initial value inspired by the semi-major axis of \citep{neuhauser2011}, leaving other orbital elements to cover wider values in the parameter space, as set by standard values from ORVARA. The specific priors and initial values set are detailed in Table~\ref{priors}. Further orbital constraints and results are analyzed and presented in section \ref{sec:orbitalcutoff}.


\begin{table}[!htb]
\begin{minipage}{0.5\textwidth}
\centering
\caption{Priors and initial  and distribution of parameters for $\eta$ Tel system astrometrical fitting$^{a}$.}
\label{priors}
\begin{tabular}{l|cc}
            Priors & $\eta$ Tel A           & $\eta$ Tel B          \\
                 \hline
Mass \, (\MSun) & 2.09$\pm$0.03      & 0.045$\pm$0.014   \\
Parallax (mas)     & 20.603$\pm$0.099          & 20.603$\pm$0.099         \\
RA p.m. (mas/yr)       & 25.689$\pm$0.006        & 25.689$\pm$0.006       \\
DEC p.m. (mas/yr)     & -82.807$\pm$0.006         & -82.807$\pm$0.006         \\
\hline
Initial distribution$^{b}$ & & \\
\hline
\(a\) (au)        & ---        & 220 $^{+300}_{-220}$          \\
$\sqrt{e}*sin(\omega)$     & --- &  0.4$\pm$0.3 \\
$\sqrt{e}*cos(\omega)$      & --- & 0.4$\pm$0.3 \\
\(i\) (radians)     & --- & 1.57$\pm$1.57 \\
$\Omega$ (radians)     & --- & 3.2$\pm$2.2 \\
$l$ (radians)     & --- & 0.8$\pm$0.5 \\
\end{tabular}
\\
\raggedright \footnotesize
$^{a}$ Unphysical values such as negative values for masses or semi-major axis are automatically excluded from the code.
\\
\raggedright \footnotesize $^{b}$ The initial distributions involve a lognormal distribution for the semi-major axis, while all other orbital values follow a normal distribution.
\end{minipage}
\end{table}

\section{Discussion}

\subsection{Orbital fitting constraints}
\label{sec:orbitalcutoff}
From the results of the MCMC simulation, we excluded the orbits that may cause instability in the system. In order to prevent disruption of the debris disk by the close-by passage of $\eta$ Tel B, and considering that the masses of the star and companion are well-constrained by the MCMC simulations, we can delimit the possible orbits for $\eta$ Tel B by inspecting the brown dwarf-disk interaction. Before applying any constraint, we obtained as values the mass of $\eta$ Tel A ($m_{*} = 2.09 M_{\odot}$) and mass of $\eta$ Tel B ($m_{comp} = 48.10 \MJup$) from the original ORVARA simulation.

In this context, it is useful to introduce the concept of chaotic zone, a region in the proximity of the orbit of a planet or brown dwarf which is devoid of dust grains, since its gravitational influence sweeps out small dust particles. The chaotic zone depends on the ratio between the mass of the brown dwarf and the mass of the star, on the semi-major axis and eccentricity of the orbit. Given the outer disk's position, it is established that this zone cannot extend below 24 au. As $\eta$ Tel B is positioned outside the disk, we employed the procedure outlined in \cite{Lazzoni2018} to compute the chaotic zone's extension for each orbit during the pericenter passage. Specifically, we utilized equation 10 for cases where the eccentricity is less than a critical value ($e_{crit}$), or equation 12 for cases where it exceeds $e_{crit}$. Here, $e_{crit}$ is determined as 0.21$\mu^{3/7}$, equivalent to 0.022 ($\mu$ is the ratio between the mass of the companion and the mass of the star). In the end, we discarded 326674 orbits from an original number of 500000.

Furthermore, another constraint arises when determining that the pericenter of the brown dwarf (BD) cannot reach the disk. Consequently, considering the semi-major axis of the companion established by the MCMC simulations, ($a_{bd}\footnote[1]{The value of the semi-major axis of 178 au was the one retrieved before the exclusion of orbits. The final value is 218 au, as presented in Table \ref{orvararesults} and Fig. \ref{cornerplot}.} = 178~au$), we can use the following expression to discard highly eccentric orbits: 

\begin{equation}
\centering
a_{bd}(1-e_{max}) > 24 ~au
\end{equation}
where $e_{max}$ is the maximum eccentricity allowed for the BD. Consequently, we obtained a maximum eccentricity of 0.865 and, therefore, we discarded an additional 139 orbits.
Another constraint can be applied if we consider that the $\eta$ Tel system has a wide comoving object, as mentioned previously in \citet{neuhauser2011}. HD 181327 is an F5.5 V star, also a bona-fide member of the $\beta$ Pic moving group, at a separation of $\sim$7 arcmin (20066 au) \citep{holmberg2009,neuhauser2011,gaiadr3}. 
The small proper motion difference in Gaia DR3, corresponding to a velocity difference on the plane of the sky of 370 m/s, is compatible with a bound object. The nominal RV difference derived in \citealp{gaiadr3, 2021A&A...645A..30Z}, is instead larger than the maximum expected one for a bound object ($\sim$400 m/s).  
However, it is well possible that the published RV errors are underestimated for a star with an extremely fast $v \sin i$ such as $\eta$ Tel. There could also be contributions by additional objects, although both the Gaia RUWE (Gaia Renormalized Unit Weight Error) and the analysis of homogeneous RV time series do not indicate the presence of close companions \citep{lagrange2009,grandjean2020}.
Therefore, we consider plausible, although not fully confirmed, that HD 181327 is bound to the $\eta$ Tel system.

\begin{figure}[!htb]
\begin{center}        \includegraphics[width=0.45\textwidth]{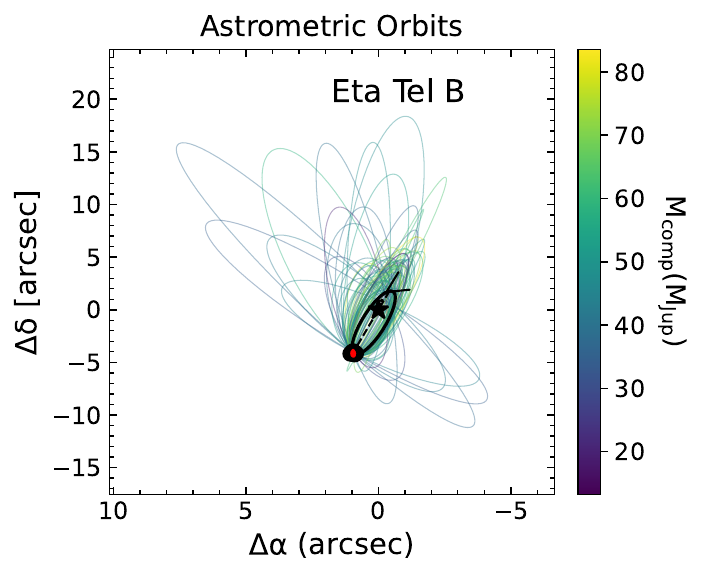}
        \caption{87 randomly selected possible orbits calculated by the Orvara fitting,  after excluding orbits based on constraints described in Section {\color{red}4.1}. The black star indicates the position of $\eta$ Tel A. The best-fit orbit is shown in black.}
    \label{orbits}
\end{center}
\end{figure}

To ensure that HD 181327's presence in the system would not affect the long-term stability of $\eta$ Tel B, we implemented equation 1 of \citep{holman1999}. Therefore, we could exclude orbits where the critical semi-major axis is greater than the periastron of the brown dwarf. To proceed, we used the mass of the perturber as 1.3 $M_{\odot}$ \citep{desidera2021,desidera2021catalogue}. Considering the eccentricities of the binaries as 0, a perturber with a similar mass would only impose constraints with a separation less than 13\farcs25 (656 au) from $\eta$ Tel A. Otherwise, for HD 181327 to act as a perturber at its separation from the $\eta$ Tel system, the eccentricities between HD 181327 and $\eta$ Tel should be greater than 0.947. Therefore, we choose not to exclude any orbit using the external perturber criteria.
Following, the results of the simulation after the cut-off are shown as the corner plot in Fig.~\ref{cornerplot} and also summarized in Table~\ref{orvararesults}. The values obtained depict a pericenter of 2\farcs9 (144 au) and an apocenter of 5\farcs9 (292 au). Despite two decades of observations of $\eta$ Tel B, the coverage only spans a fraction of the wide orbit of the companion, with a total $\Delta$PA of approximately $\sim$2$^{\circ}$. This accounts for less than 1\% of the orbit, assuming a face-on and circular orbit. Consequently, constraining the semi-major axis and eccentricity proves challenging. Nevertheless, the MCMC effectively determines the inclination, yielding an orbit that is nearly edge-on and almost co-planar with the debris disk \citep{2009A&A...493..299S}. 
Conversely, a variety of possible orbits can adequately fit the data, as illustrated in Fig.~\ref{orbits}.

\begin{table}[!htb]
\caption{Best orbital-fitting parameters for $\eta$ Tel B calculated from the Orvara orbital characterization.}
\label{orvararesults}
\begin{tabular}{l|l}
Mass (\MJup)                   & ${48}_{-15}^{+15}$         \\
\(a\) (au)                       & ${218}_{-41}^{+180}$            \\
\(i\) (deg)            & ${81.9}_{-3.5}^{+3.2}$          \\
\(\Omega\) (deg)         & ${174.6}_{-7.1}^{+175}$         \\
Mean longitude (deg)         & ${184}_{-74}^{+164}$           \\
Parallax (mas)               & ${20.60}_{-0.10}^{+0.10}$   \\
Period (yrs)                 & ${2201}_{-592}^{+3224}$         \\
\(\omega\) (deg) & ${159}_{-99}^{+128}$            \\
\(e\)                 & ${0.34}_{-0.23}^{+0.26}$       \\
\(a\) (mas)         & ${4486}_{-845}^{+3704}$        \\
Reference epoch \(T0\)  (JD)                      & ${2740996}_{-107450}^{+488958}$ \\
Mass ratio                   & ${0.0219}_{-0.0068}^{+0.0069}$
\end{tabular}
\end{table}



\subsection{Possible formation scenarios}
\label{sec:formationscenario}

The coplanarity between $\eta$ Tel B and the debris disk, as described in this work, and the relative spin-alignment between the star and the debris disk \citep{hurt2023}, can offer insights into the formation scenario of the system. Stellar systems form through various mechanisms, typically categorized into three main types: 1) fragmentation of a core or filament, 2) fragmentation of a massive disk, or 3) capture and/or dynamical interactions (for a comprehensive review, including separation and formation time scales, see \citealp{offner2022}). Given the separation of the star and the companion, scenarios involving massive disk fragmentation or capture and/or ejection of $\eta$ Tel B appear more plausible.

Alternatively, core fragmentation, which occurs through direct/turbulent fragmentation (e.g., \citealp{1979ApJ...234..289B, 1997MNRAS.288.1060B}) or rotational fragmentation (e.g., \citealp{larson1972,1994MNRAS.269..837B,1994MNRAS.269L..45B,1994MNRAS.271..999B,1997MNRAS.289..497B}), followed by inward migration, is also a conceivable formation scenario. Binaries formed through direct fragmentation, with well-separated cores, may exhibit uncorrelated angular momenta between the objects \citep{2016ApJ...827L..11O, 2018MNRAS.475.5618B, 2019ApJ...887..232L, 2000MNRAS.314...33B}. In contrast, stars and substellar objects formed through rotational fragmentation within the same plane tend to display preferentially spin-aligned and coplanar systems \citep{2016ApJ...827L..11O, 2018MNRAS.475.5618B}. In such a case, a rotational fragmentation followed by inward migration becomes more probable.

Moreover, the eccentricity of the system plays a crucial role in inferring the formation scenario. If we consider the orbital fitting results where highly eccentric orbits are permitted, the preferred hypothesis for eccentricity enhancement would be recent dynamical interactions.  If so, $\eta$ Tel B did not reach near the debris disk of the star in the last few pericenter approximations. This scenario is highly improbable, given the short best-fitting orbital period in such cases ($\sim{1623}$ years). It is more likely that $\eta$ Tel B has a low eccentric orbit, and the assumptions for excluding highly eccentric orbits may be the most suitable approach. Consequently, the system exhibits quasi-coplanarity and low eccentricity of the companion. Therefore, we tentatively suggest that the preferred formation scenarios for $\eta$ Tel involve either the fragmentation of a massive disk with slow or no inwards migration or rotational fragmentation of a core with faster inwards migration. Long-term RV and astrometry monitoring of the star and the companion, along with multiwavelength observations of the system, could be useful to discard a capture or ejection scenario.


\subsection{The close surroundings of $\eta$ Tel B }
\label{sec:sathunt}

In our Solar System, planets and small-sized bodies are often surrounded by satellites and dust rings or disk-like features (see e.g. \citealp{alibert2005}). For instance, there are approximately 200 natural satellites in the Solar System, most of which orbit giant planets. This raises the possibility of similar objects existing around substellar companions, although such discoveries have not yet been confirmed. Satellites or circumplanetary disks in these environments, if discovered, could provide valuable insights into their formation mechanisms. These mechanisms may include gravitational instability (\citealp{boss1997} and references therein), core-accretion (\citealp{pollack1996} and references therein), or capture and/or orbital crossing, which can lead to satellite companions with specific mass ratios and orbits. For instance, less massive exomoons are likely to form within a circumplanetary disk (CPD), as observed with the Galilean moons (see, e.g., \citealp{canupeward2002}). Conversely, massive companion+satellite candidate systems likely form via orbital crossing+capture \citep{ochiai2014, lazzoni2024}. Furthermore, hydrodynamical simulations have shown that CPDs in the core-accretion scenario are eight times less massive and one order of magnitude hotter than those formed by gravitational instability \citep{szu2017}. Consequently, the characterization of exosatellites can be used to distinguish between these formation scenarios.

\begin{figure*}[!htb]
\begin{center}
    \includegraphics[width=0.9\textwidth]{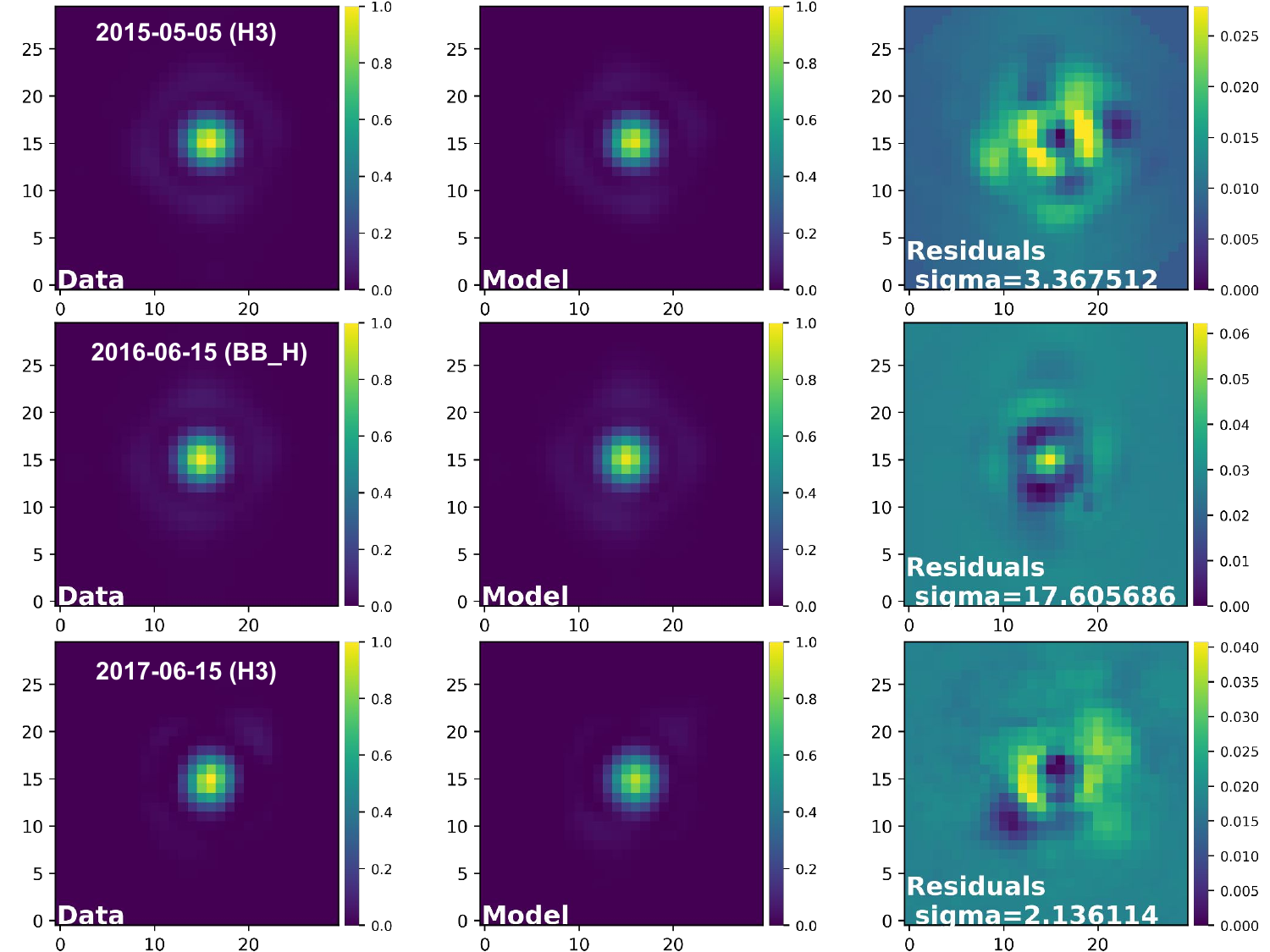}
    \caption{Analysis of the close surroundings of $\eta$ Tel B to look for satellites and/or CPDs. In each row, the first panel shows the ADI image zoomed on $\eta$ Tel B, the second panel shows the model, and the last one the residuals from the subtraction of the two. The three IRDIS epochs are shown. The residuals are expressed in counts. All the panels were normalized to the peak of the central PSF.}
    \label{fig:residuals}
\end{center}
\end{figure*}


To analyze whether a satellite candidate or a CPD is present around the companion, we proceeded as follows: to compensate for the self-subtraction effect induced by post-processing, we forward-modeled the companion per frame based on the observations and the PSF model. With the pre-processed data, the radial distance, position angle, and flux were obtained from the NEGFC approach (refer to Section \ref{sec:astrometry_negfc}). The off-axis PSF extracted before the sequence was positioned and flux-normalized with these parameters in an empty frame, creating what we refer to as ``model''. Following, for each frame, the model was subtracted from the data, producing a residual image. we collapsed the model and residual cubes per night using ADI or PCA+ADI. We compare the ADI post-processed data with the collapsed model and residual cubes per night in Fig. \ref{fig:residuals}). The quality of the residuals is strongly related to the number of frames and quality of the night, showing a clearer result for the epoch 2016-06-15, where other epochs are more affected by systematics. Still, the residuals do not show any clear signal of a satellite or other nearby structure, imposing a threshold of detections at the contrast obtained with the NEGFC technique. 

To generate contrast curves around $\eta$ Tel B, we implemented a methodology similar to the one outlined in \citet{lazzoni2020}. The steps can be summarized as follows. Successive annuli, each centered on $\eta$ Tel B and with a width equal to 1 FWHM, were chosen up to the Hill radius of the brown dwarf. The contrast at each radial position is computed as five times the standard deviation inside the annulus divided by the peak of the star ($\eta$ Tel A).

\begin{figure}[!h]
\begin{center}
    \includegraphics[width=0.45\textwidth]{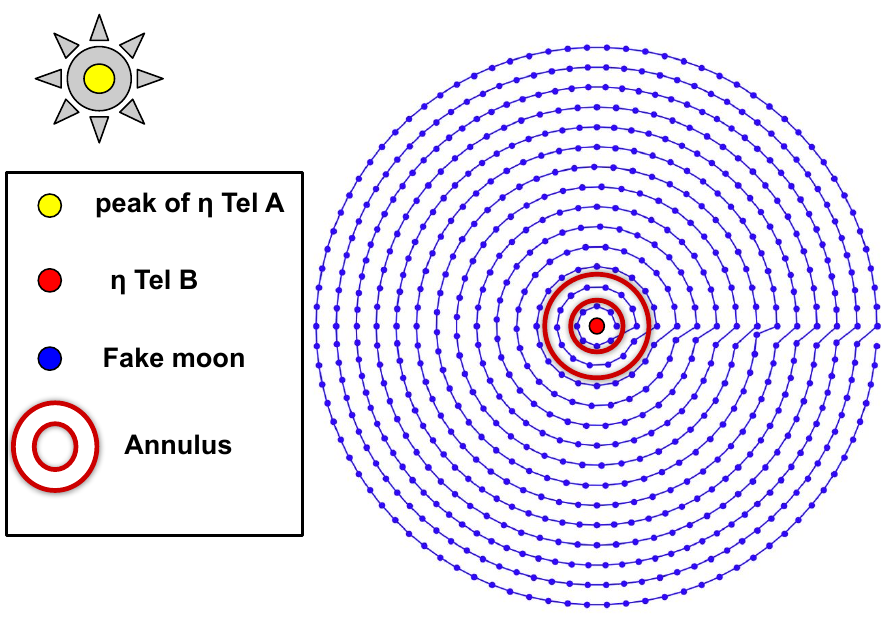}
    \caption{Schematic representation of the placement of putative fake satellites around $\eta$ Tel B to calculate the contrast curves. In this illustration, just one annulus is represented.}
    \label{illustration_cc}
\end{center}
\end{figure}

For each annulus, we injected fake companions at various position angles, and their fluxes were determined after applying the NEGFC and PCA post-processing techniques. The ratio between the retrieved and injected flux provides the throughput value. A mean throughput is then calculated for each annulus and multiplied by the contrast at each separation. As a final step, we adjusted the contrast for small sample statistics, following the discussion presented in \citet{mawet2014}. A schematic representation of the steps used to calculate the contrast curves is illustrated in Fig.~\ref{illustration_cc}.

\begin{figure}[!h]
\begin{center}    
\includegraphics[width=0.45\textwidth]{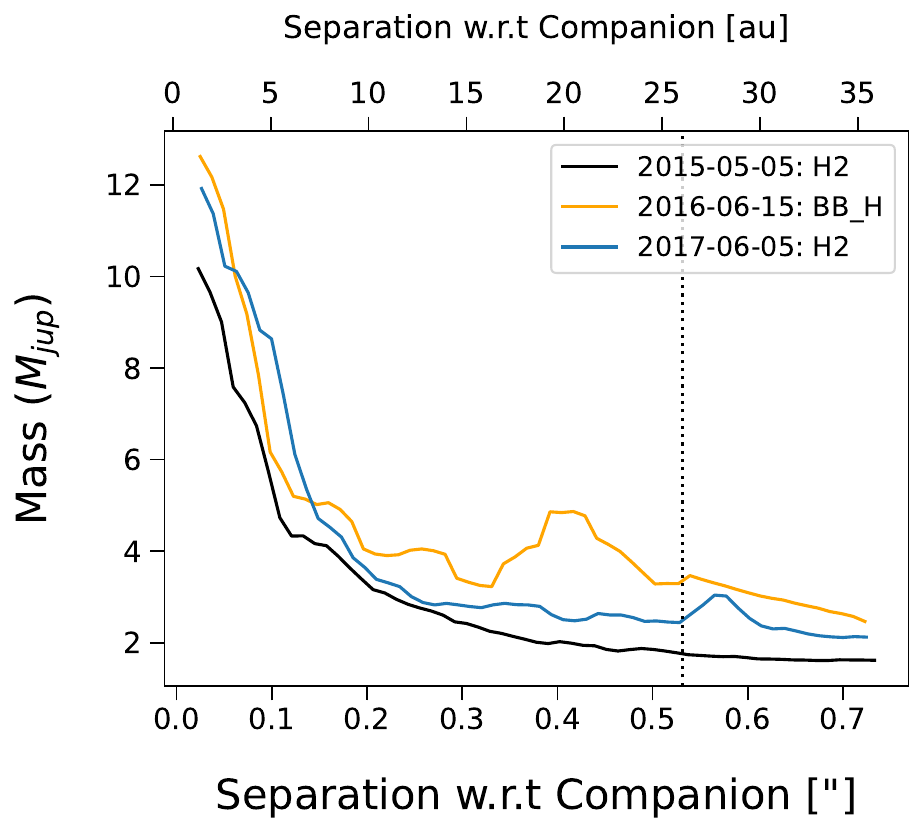}
    \caption{
    Contrast curves with respect to the star $\eta$ Tel A, centered on the companion as depicted in Fig. \ref{illustration_cc}. The three contrast curves, corresponding to different epochs, are showcased.
    The vertical dashed line indicates the Hill radius of $\eta$ Tel B under the assumption of an orbit with an eccentricity of $e$=0.34}.
\label{contrast_curves}
\end{center}
\end{figure}

Finally, the contrasts were converted into mass constraints using the ATMO 2020 evolutionary models \citep{Phillips}. We utilized an estimated age for the system of 18 Myr \citep{2020A&A...642A.179M} and a distance of 49.5 pc \citep{gaiadr3}. The best contrast curves for each data set are depicted in Fig.~\ref{contrast_curves}. We can discard the detection of satellites around the brown dwarf with masses between 3 and 1.6 \MJup in the range of distances  $[10,33]$ au. Such massive objects, if present, would likely be the result of capture or trapping through tidal interactions \citep{lazzoni2024} or formation in situ via gravitational instability or direct collapsing. However, we cannot exclude the presence of closer-in and/or less massive objects which, for example, could have formed within a CPD via core accretion.
Moreover, we can exclude the presence of an extended CPD from the shape and luminosity of the residuals around the companion.


\section{Conclusions}

\label{sec:sum}
In this paper, we present the most recent photometric and astrometric characterization of $\eta$ Tel B, a brown dwarf situated at an approximate separation of $\sim$4$\farcs$2 from its host star. The observations of this system were conducted using SPHERE/IRDIS H2H3 and BB\_H filters, spanning three epochs across three consecutive years (2015-2017).

To robustly establish astrometry and photometry for the sub-stellar companion, we employed the NEGFC customized routine presented in \citep{lazzoni2020}. This approach was applied to each frame of the scientific datacube, as opposed to solely in the post-processed ADI image, as is conventionally practiced. Photometric results were derived by considering the median of each set of parameters per observation. Photometric errors were determined based on the standard deviation of the fluxes obtained per night. Astrometrical results (separation of 4.218$\pm$0.004 arcsecs and position angle of 167.3$\pm$0.2 degrees) were made taking into account not only the NEGFC approach but also incorporated an analysis using 2D Gaussian fitting and position of the peak intensity of the companion. This analysis was made for both the frame-by-frame approach and the median-collapsed ADI image. The final values represent the median of each approach. In addition, for the uncertainties, systematic and statistical uncertainties, akin to the methodology employed by \citet{wertz2017}, were employed. The separation reached depicts a 4-70 times improvement in precision in comparison with previous NACO observations described and observed by \citet{neuhauser2011}, and $\sim$2 times better precision than the L$^{\prime}$ NACO observations from \citet{rameau2013}. 

Furthermore, we conducted a comprehensive orbital characterization by compiling previous astrometric data, the new SPHERE data, and the Hipparcos-Gaia acceleration catalog, resulting in an orbital coverage spanning approximately 19 years. The findings indicate a well-characterized companion mass of 48$\pm$15 \MJup, with the best-fitting orbit demonstrating near-edge-on orientation ($i$=81.9 degrees) and low eccentricity ($e=0.34$) when excluding orbits that can disrupt the debris disk around the star. However, it is essential to note that our observations did not encompass a significant portion of the entire orbit, leading to elevated uncertainties regarding the companion's orbital shape, period, eccentricity, and semi-major axis. We also highlight that due to the low orbital coverage, the orbital fitting presented represents a plausible family of orbits, and the orbital values listed in Table~\ref{orvararesults} must be taken as an example of a possible orbit. Further follow-ups will better constrain the orbit of the BD in the future. Also, based on the high $v sin i$ of $\eta$ Tel A, and the possible uncertainties on its measurements, we consider it plausible that HD 181327, at a separation of 20066 au, is bound to the system. 

A brief discussion about possible formation scenarios has been conducted. The coplanarity between $\eta$ Tel B and its debris disk, along with the relative spin alignment between the star and the debris disk, provides valuable insights into the system's formation. Stellar system formation scenarios were categorized into three main types: fragmentation of a core or filament, fragmentation of a massive disk, or dynamical interactions. Based on the separation of the star and companion, the likelihood of massive disk fragmentation or capture or ejection of $\eta$ Tel B was highlighted as more plausible. Alternatively, the system's low eccentricity also allows for rotational fragmentation from a core scenario, if followed by fast inward migration. In summary, we tentatively propose the preferred formation involves either massive disk fragmentation with slow or no inward migration or rotational fragmentation of a core with faster inward migration, whereas continuous monitoring and multi-wavelength observations are crucial for refining these conclusions and further understanding the system's dynamics.

Lastly, a meticulous analysis of the companion's surroundings was undertaken by subtracting the companion's signal using instrument-response models. No clear signal of a substructure or satellite was seen. We conclude the systematics heavily affected the residuals. From the contrast curves generated in the regions surrounding the companion, we can discard satellites down to 3 and 1.6 \MJup in the range of distances [10, 33] au, setting an upper limit on gravitational instability binary pairs or massive objects captured through tidal interactions at wider separations.

Future observations of the system with the next generation of high-contrast imagers mounted on space telescopes, likewise, JWST (James Webb Space Telescope) will further constrain the orbital analysis of the brown dwarf companion and allow for deeper contrast around the central star and the companion. This advancement will enable deeper contrast observations around both the central star and its companion. In the specific case of RV monitoring for detecting objects at closer separations, instruments like CRIRES+ (Cryogenic high-resolution InfraRed Echelle Spectrograph) or HiRISE (High Resolution Imaging Science Experiment),  are suitable. Possible additional companions to the central star or the brown dwarf may be detected in the future.

\begin{acknowledgements}
The authors thank the anonymous referee for the fruitful comments and suggestions. The authors acknowledge support from ANID -- Millennium Science Initiative Program -- Center Code NCN2021\_080. P.H.N. acknowledges support from the ANID Doctorado Nacional grant 21221084 from the government of Chile. T.B. acknowledges financial support from the FONDECYT postdoctorado project number 3230470. SD acknowledges support by the PRIN-INAF 2019 ``Planetary systems at young ages (PLATEA)''. S.P. acknowledges support from FONDECYT Regular grant 1231663. This work has made use of the High Contrast Data Centre, jointly operated by OSUG/IPAG (Grenoble), PYTHEAS/LAM/CeSAM (Marseille), OCA/Lagrange (Nice), Observatoire de Paris/LESIA (Paris), and Observatoire de Lyon/CRAL, and supported by a grant from Labex OSUG@2020 (Investissements d’avenir – ANR10 LABX56). 
\end{acknowledgements}

\bibliographystyle{aa}
\bibliography{aa}

\begin{appendix}

\section{Cornerplot of the orbital fitting}
\begin{figure}[htb!]
    \begin{minipage}[t]{1\textwidth} 
        \begin{tabular}[t]{ @{} r @{} }
        \includegraphics[width = 1\linewidth]{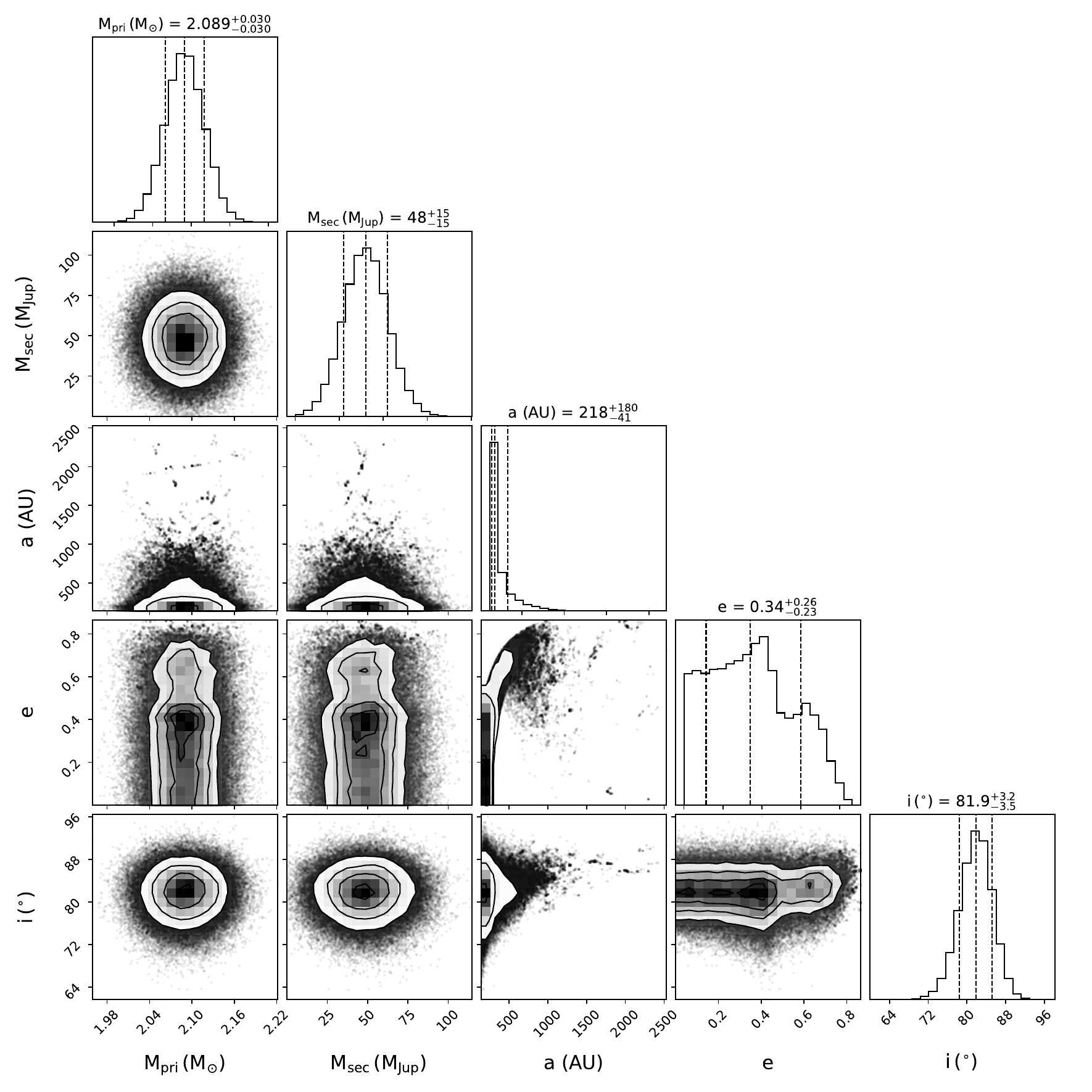} \\

       \end{tabular}
               \caption{Corner plot showing the results of the orbital fitting performed by Orvara, after excluding orbits based on constraints. The best-fit values are also reported in Table~\ref{orvararesults}.}
    \label{cornerplot}
    \end{minipage}
\end{figure}





\end{appendix}
\end{document}